\documentclass[10pt,a4paper]{article}

\usepackage{amsmath,amsgen,latexsym}
\usepackage{amstext,amssymb,amsfonts,latexsym}
\usepackage{theorem}
\usepackage{pifont}

 \newcommand{\bs}{\bigskip}
 \newcommand{\ms}{\medskip}
 \newcommand{\n}{\noindent}
 \newcommand{\s}{\smallskip}
 \newcommand{\hs}[1]{\hspace*{ #1 mm}}
 \newcommand{\vs}[1]{\vspace*{ #1 mm}}

 \newcommand{\nat}{\mathbb{N}}
 
 \newcommand{\integer}{\mathbb{Z}}

 \newcommand{\complex}{\mathbb{C}}
 
 \newcommand{\ptcomplex}{\tilde{\mathbb{C}}}
 
 \newcommand{\algebraic}{\mathbb{A}}

 \newcommand{\ie}{\textrm{i.e.},\hspace*{2mm}}
 \newcommand{\eg}{\textrm{e.g.},\hspace*{2mm}}
 
 \newcommand{\etalc}{\textrm{et al.}}

 \newcommand{\p}{\mathrm{P}}
 \newcommand{\np}{\mathrm{NP}}

 \newcommand{\am}{\mathrm{AM}}

 \newcommand{\pspace}{\mathrm{PSPACE}}
 
 \newcommand{\e}{\mathrm{E}}

 \newcommand{\nexp}{\mathrm{NEXP}}
 \newcommand{\eespace}{\mathrm{EESPACE}}

 \newcommand{\ip}{\mathrm{IP}}
 \newcommand{\twoip}{\mathrm{2IP}}

 \def\bbox{\vrule height6pt width6pt depth1pt}

\theoremstyle{plain}
\theoremheaderfont{\bfseries}
 \newtheorem{theorem}{Theorem}[section]
 \newtheorem{lemma}[theorem]{Lemma}
 \newtheorem{proposition}[theorem]{Proposition}
 \newtheorem{corollary}[theorem]{Corollary}

 {\theorembodyfont{\rmfamily}
 }
 {\theorembodyfont{\rmfamily} }
 {\theorembodyfont{\rmfamily} }

 \newenvironment{proofof}[1]{\vspace*{5mm} \par \noindent
         {\bf Proof of #1.\hs{2}}}{\hfill$\Box$ \vspace*{3mm}}

 \newcommand{\ceilings}[1]{\lceil #1 \rceil}
 
 \newcommand{\pair}[1]{\langle #1 \rangle}

 \newcommand{\dtime}[1]{{\mathrm{DTIME}}(#1)}
 \newcommand{\ntime}[1]{{\mathrm{NTIME}}(#1)}

 \newcommand{\dspace}[1]{{\mathrm{DSPACE}}(#1)}

 \newcommand{\qip}{\mathrm{QIP}}
 \newcommand{\twoqip}{\mathrm{2QIP}}
 
 \newcommand{\qmip}{\mathrm{QMIP}}
 \newcommand{\mip}{\mathrm{MIP}}

\newif\ifnotesw\noteswtrue
   {\ifnotesw\marginpar[\hfill\(\top\)]{\(\top\)}\fi}%
      {\ifnotesw\marginpar[\hfill\(\bot\)]{\(\bot\)}\fi}
      
\newcommand{\mnote}[1]%
   {\ifnotesw\marginpar%
	  [{\scriptsize\begin{minipage}[t]{\marginparwidth}
	  \raggedleft#1%
		  \end{minipage}}]%
	  {\scriptsize\begin{minipage}[t]{\marginparwidth}
	  \raggedright#1%
		  \end{minipage}}%
    \fi}

\newcommand{\ignore}[1]{}

\newcommand{\cent}{{|}\!\!\mathrm{c}}
\newcommand{\track}[2]{\left[ \begin{smallmatrix} %
     #1 \\ #2 \end{smallmatrix} \right]}

 \newcommand{\polytime}{poly\mbox{-}time}

 \newcommand{\cfl}{\mathrm{CFL}}

 \newcommand{\qubit}[1]{| #1 \rangle}
 \newcommand{\bigqubit}[1]{\left| #1 \right\rangle}
 \newcommand{\bra}[1]{\langle #1 |}
 \newcommand{\ket}[1]{| #1 \rangle}

 \newcommand{\measure}[2]{\langle #1 | #2 \rangle}

\begin{document}
\pagestyle{plain}
\setcounter{page}{1}


\begin{center}
{\Large {\bf Constant-Space
 Quantum Interactive Proofs  \s\\ Against Multiple Provers}}\footnote{The results of this paper were first reported at the 4th Central European Quantum Information Processing Workshop (CEQIS 2007), June 24--27, 2007, Valtice, Czech Republic.} \bs\s\\

{\sc Tomoyuki Yamakami}\footnote{Present Affiliation: Department of Information Science, University of Fukui, 3-9-1 Bunkyo, Fukui 910-8507, Japan} \end{center}
\bs\s

\begin{abstract}
\n We present upper and lower bounds of the computational complexity of the two-way communication model of multiple-prover quantum interactive proof systems whose verifiers are limited to measure-many two-way quantum finite automata. We prove that (i) the languages recognized by those multiple-prover systems running in expected polynomial time are exactly the ones in NEXP, the nondeterministic exponential-time complexity class, (ii) if we further require verifiers to be one-way quantum automata, then their associated proof systems recognize context-free languages but not beyond languages in NE, the nondeterministic linear exponential-time complexity class, and moreover, (iii) when no time bound is imposed, the proof systems become  as powerful as Turing machines. The first two results answer affirmatively an open question, posed by Nishimura and Yamakami [J. Comput. System Sci, 75, pp.255--269, 2009], of whether multiple-prover quantum interactive proof systems are more powerful than single-prover ones. Our proofs are simple and intuitive, although they heavily rely on an earlier result on multiple-prover interactive proof systems of Feige and Shamir [J. Comput. System Sci., 44, pp.259--271, 1992].

\ms

\n{\sf Keywords:} theory of computing,
formal languages,
quantum interactive proof system,
quantum finite automaton,
nondeterministic exponential time,
recursively enumerable language,
context-free language
\end{abstract}

\section{Background and Main Results}

Quantum interactive proof (QIP) systems have drawn significant attention lately \cite{JJU+11,KW00,KM03,NY04b,Wat03,Yao03}. A QIP system for a target language $L$ is a two-player game in which a series of interactions between a mighty prover and a computationally-limited verifier eventually leads the verifier to determine whether a given word belongs to this particular
language $L$ with a small margin of error.
The role of such a prover has two sides: a {\em honest} prover tries to convince the verifier to accept the word if the word is truly in the language, while a {\em dishonest} prover tries to misguide the verifier to accept it if the word is outside of the language. Such a system can be sought as an extension of the ``proof-and-verification'' characterization of $\np$ languages (where a prover nondeterministically presents a proof and a verifier deterministically examines its correctness). Recent studies on various QIP systems (including quantum Arthur-Merlin proof systems and quantum zero-knowledge proof systems) have shown their significant power in recognizing languages, compared to classical interactive proof (IP) systems. It has been important to expand the scope and depth of research on QIP systems for their better understandings. Along this line of research, this paper looks into QIP systems whose verifiers are allowed to use only constant memory space against powerful {\em multiple} provers.

In the early 1990s, Dwork and Stockmeyer \cite{DS92a} published their seminal paper on the computational complexity of weak-verifier IP systems, where verifiers behave particularly as two-way probabilistic finite automata\footnote{Although a finite automaton can store a certain type of information in the form of its tape-head location, a finite automaton is generally viewed as a model of a constant-space computer because its central processing unit uses only a constant number of inner states.} (or 2pfa's, in short). Although the verifier is weak, the corresponding IP systems are relatively powerful. A major advantage of studying such {\em constant-space}   verifiers is that we can prove certain class separations, which are impossible (at present) for polynomial-time or logarithmic-space bounded IP systems. To describe earlier results on constant-space-verifier IP systems, it is useful at this point to introduce several notations. Adapting the notations of Dwork and Stockmeyer, we denote by $\ip(2pfa)$ and $\ip(2pfa,\polytime)$ the classes of all languages that admit IP systems whose verifiers are respectively 2pfa's and expected-polynomial-time 2pfa's. Moreover, their {\em public-coin} versions (or Arthur-Merlin versions) are denoted by $\am(2pfa)$ and $\am(2pfa,\polytime)$ accordingly. Dwork and Stockmeyer proved several upper and lower bounds of those complexity classes: namely, $\mathrm{REG}\subseteq \am(2pfa,\polytime) \subsetneqq \am(2pfa) \subsetneqq \ip(2pfa,\polytime) \subseteq \ip(2pfa) \cap \pspace$, $\mathrm{2PFA}\subsetneqq \am(2pfa) \subsetneqq \p$, and $\e \subseteq \ip(2pfa) \subseteq \eespace$, where $\mathrm{REG}$ denotes the class of all regular languages, $\e=\dtime{2^{O(n)}}$, and $\eespace = \dspace{2^{2^{O(n)}}}$.
For a quick overview of this field, see Condon's comprehensive survey \cite{Con93} on space-bounded IP systems.

Motivated by the aforementioned work of Dwork and Stockmeyer, Nishimura and Yamakami \cite{NY04b} (for its journal versions, see \cite{NY09,NY13}) studied in 2004 weak-verifier QIP systems in which verifiers are variants of {\em (measure-many) quantum finite automata} (or qfa's, in short). Such qfa's were initially introduced as a direct extension of pfa's and they are still of theoretical interest because of their simplicity as a quantum mechanical model of computation. Nishimura and Yamakami
obtained the following preliminary containments and separations. As in \cite{NY04b,NY09,NY13}, let $\qip(\pair{restrictions})$ denote the class of all languages recognized with small error probability by certain QIP systems that satisfy all restrictions specified by $\pair{restrictions}$. When verifiers are limited to one-way quantum finite automata (or 1qfa's), the corresponding complexity class $\qip(1qfa)$ equals $\mathrm{REG}$. In contrast,
if verifiers are two-way quantum finite automata (or 2qfa's),
then it holds that $\mathrm{REG}\subsetneqq \qip(2qfa,\polytime)\nsubseteq \am(2qfa)$ and $\qip(2qfa,\polytime)\subseteq \np$, provided that  verifier's amplitudes are restricted to polynomial-time approximable  complex numbers. Nevertheless, no inclusion relationship is known between classical-prover QIP systems and quantum-prover QIP systems mainly because of the constant  constraint of the verifier's memory space.

The aforementioned results of Dwork and Stockmeyer \cite{DS92a} and of Nishimura and Yamakami \cite{NY04b,NY09,NY13} concern single-prover systems. A natural extension of such a proof system is a {\em multiple-prover system}, in which a single verifier communicates with two or more provers through separate communication channels.
Shortly after the work of Dwork and Stockmeyer, Feige and Shamir \cite{FS92} introduced a weak-verifier model of multiple-prover interactive proof (MIP) systems. For convenience, we write $\twoip(2pfa)$ and $\twoip(2pfa,\polytime)$ respectively for the complexity classes induced by 2-prover IP systems (or 2IP systems, in short) with 2pfa-verifiers and with expected-polynomial-time 2pfa-verifiers. Such systems turn out to possess an enormous power; namely, $\twoip(2pfa) = \mathrm{RE}$ and $\twoip(2pfa,\polytime) =\nexp$, where $\mathrm{RE}$ is the class of all {\em recursively enumerable languages} and $\nexp$ denotes $\mathrm{NTIME}(2^{n^{O(1)}})$. The latter result is built upon a result of Babai, Fortnow, and Lund \cite{BFL91}, who proved that polynomial-time bounded MIP systems are exactly as powerful as the complexity class $\nexp$.

Returning to the case of QIP systems, Kobayashi and Matsumoto \cite{KM03} discussed the computational complexity of polynomial time-bounded multiple-prover quantum interactive proof (QMIP) systems and they showed that such systems are also as powerful as the class $\nexp$. This result establishes a bridge between classical and quantum computations based on interactive proof models of computation.
Similarly to  $\twoip$ systems of Feige and Shamir, we  formulate and study a multiple-prover model of constant-space-verifier QIP systems. Our model naturally expands the aforementioned single-prover model of Nishimura and Yamakami by allowing more than one prover to interact with a single verifier.
We use the notation $\qmip(\pair{restrictions})$ to indicate the class of all languages that admit multiple-prover QIP systems with the restrictions specified by $\pair{restrictions}$.  See Section \ref{sec:basics} for the formal definition of $\qmip(\pair{restrictions})$. In their paper
\cite[Section 6]{NY09}, Nishimura and Yamakami posed a fundamental question of whether multiple-prover QIP systems are more powerful than single-prover QIP systems.

Built upon the earlier results of Feige and Shamir \cite{FS92} and of Kobayashi and Matsumoto \cite{KM03}, we give a complete characterization of QMIP systems with 2qfa-verifiers who run
in expected polynomial time by showing that such systems are computationally
equivalent to classical MIP systems of Feige and Shamir; therefore,
they match the complexity class $\mathrm{NEXP}$ in computational power.
A main technical achievement here is an extension of the cryptographic trick of Feige and Shamir against quantum provers. By removing the runtime bound of 2qfa-verifiers, we establish another bridge between 2qfa-verifier 2QIP systems and the class $\mathrm{RE}$. In the case of 1qfa-verifiers, we also obtain reasonable upper and lower bounds of the computational complexity of corresponding QMIP systems.
An important consequence of those results is the class separations between $\qip(2qfa,\polytime)$ and $\qmip(2qfa,\polytime)$ and between $\qip(1qfa)$ and $\qmip(1qfa)$ {\em without any unproven assumption} if all amplitudes are  polynomial-time approximable complex numbers. We thus completely solve the aforementioned open question of Nishimura and Yamakami. In comparison, as far as we know, it is not yet known in a classical case that $\ip(2pfa,\polytime)\neq \mip(2pfa,\polytime)$. Therefore, our results  exemplify a significant advantage of quantum computation over its classical counterpart.

\section{Weak-Verifier QMIP Systems}\label{sec:basics}

To help the reader go through this paper, we briefly provide a description of multiple-prover quantum interactive proof (QMIP) systems whose verifiers are particularly quantum finite automata. Our QMIP systems are a natural extension of single-prover quantum interactive proof (QIP) systems introduced by Nishimura and Yamakami \cite{NY04b} (cf.~\cite{NY09,NY13}), provided that multiple provers share no common information. Notice that, throughout this paper, we shall treat  single-prover systems as a special case of multiple-prover systems.

\subsection{Fundamental Notions and Notation}

Let $\nat$ be the set of all {\em natural numbers} (that is, nonnegative integers) and set $\nat^{+}=\nat - \{0\}$. Likewise, let $\integer$,  $\complex$, and $\algebraic$ be respectively the sets of all {\em integers},  of all {\em complex numbers}, and of all {\em algebraic complex numbers}.  Moreover, $\ptcomplex$ denotes the set of all {\em polynomial-time approximable complex numbers} (\ie both real and imaginary parts are deterministically approximated to within $2^{-n}$ in time polynomial with respect to an  approximation parameter $n$ given as input). Notice that $\algebraic\subseteq \ptcomplex\subseteq \complex$ holds.
For any two numbers $m,n\in\integer$ with $m\leq n$, the {\em integer interval}  $\{m,m+1,m+2,\ldots,n\}$ is denoted by $[m,n]_{\integer}$.
For any finite set $Q$, $|Q|$ denotes the {\em cardinality} of $Q$.

Generally, we use the notation $\Sigma$ for a finite nonempty input alphabet (not necessarily limited to $\{0,1\}$). In contrast with $\Sigma^*$, the notation $\Sigma^{\infty}$ denotes the set of all {\em infinite sequences} of symbols in $\Sigma$.
For our convenience, we use a special ``track'' notation of \cite{TYL04}. For any two tape symbols $a$ and $b$, the notation $\track{a}{b}$ means that the tape is split into two tracks and the symbol $a$ is written in its upper track and $b$ is in its lower track of the same tape cell. For any two strings $x=x_1x_2\cdots x_n$ and $y=y_1y_2\cdots y_n$, $\track{x}{y}$ denotes the sequence $\track{x_1}{y_1}\track{x_2}{y_2}\cdots \track{x_n}{y_n}$. In addition, when $|x|\neq |y|$ in the notation $\track{x}{y}$, we automatically pad extra blank symbols $\#$ to the end of the {\em shorter} string
between $x$ and $y$.
For instance, $\track{001}{10}$ means $\track{0}{1}\track{0}{0}\track{1}{\#}$.
We also use standard multi-tape off-line models of deterministic Turing machines (DTMs) and of nondeterministic Turing machines (NTMs). A tape is called a {\em read-once tape} if, whenever its tape head reads a symbol from the tape, the head should move to the right; thus, the tape head cannot access the same tape cell again.

Notationally, we denote by $\mathrm{REG}$  (resp., $\mathrm{CFL}$) the collection of all {\em regular (resp., context-free) languages}, and by $\mathrm{RE}$  the class of all {\em recursively enumerable  languages}.
Other important complexity classes to mention include $\mathrm{NE}$ ($=\ntime{2^{O(n)}}$) and $\nexp$ ($=\ntime{2^{n^{O(1)}}}$).

Throughout this paper, we use Dirac's notation $\ket{\phi}$ to express a {\em quantum state}, which is a vector in a finite-dimensional Hilbert space (occasionally, we will consider an infinite-dimensional Hilbert space). We write $\bra{\phi}$ for the {\em conjugate transpose} of $\ket{\phi}$. The {\em norm} of a quantum state $\ket{\phi}$ is given as $\|\ket{\phi}\|=\sqrt{\measure{\phi}{\phi}}$. To make this paper concise, we assume the reader's familiarity with the basics of quantum information and computation (see, \eg the textbook of Nielsen and Chuang \cite{NC00}).

\subsection{Quantum Interactive Proof Systems}

Here, we give a brief description of our multiple-prover QIP systems with weak verifiers. For a more concrete definition of (single-prover) QIP systems, refer to \cite{NY09,NY13}. Let $k$ be any number in $\nat^{+}$ and fix arbitrarily an  input length $n$ in $\nat$.
A {\em $k$-prover QIP system} (or $k\qip$, in short) $(P_1,P_2,\ldots,P_k,V)$ consists of a 2qfa-verifier $V$ and $k$ provers $P_1,P_2,\ldots,P_k$. The 2qfa-verifier $V$ has a read-only input tape, which has two special endmarkers
$\cent$ (left endmarker) and $\$$ (right endmarker) and all tape cells are  indexed by integers between $0$ and $n+1$, including these two endmarkers. Conventionally, the input tape is {\em circular} (namely, the $n+2$nd cell is the same as the $0$th cell). See \cite{KW97} for a convention of such  circular tapes.
We reserve $\#$ to denote the designated ``blank'' tape symbol, different from other symbols. For each index $i\in[1,k]_{\integer}$, the verifier $V$ communicates with the $i$th prover $P_i$ through the $i$th {\em communication cell}, which holds a symbol chosen from the $i$th communication alphabet $\Gamma_i$ containing also the blank symbol $\#$.
A single {\em move} of $V$ is completely specified
as a single application of its {\em transition function}
$\delta_{V}: Q\times \Sigma\times \Gamma_1\times \cdots\times \Gamma_k \rightarrow \complex^{Q\times \{0,\pm1\} \times \Gamma_1\times \cdots\times \Gamma_k}$, where $Q$ is a finite set of $V$'s inner states and $\{0,\pm1\}$ is the set of the tape head's directions.\footnote{The head direction $+1$ (resp., $-1$) indicates that the head moves rightward (resp., leftward) and also the direction $0$ means that the head stays still.}
After each move, $V$ applies a {\em projection measurement} to determine whether it is in halting inner states (that is, either accepting or rejecting inner states). Initially, the verifier is in its initial inner state $\qubit{q_0}$ with its tape head scanning the left endmarker $\cent$ on the $0$th cell and all the communication cells contain only $\#$s. The verifier always makes the first move.

Each prover $P_i$ has its own read/write private tape, on which it uses tape  symbols drawn from the $i$th private-tape alphabet $\Delta_i$, including
the blank symbol $\#$. Initially, the private tape of each prover consists of all blanks. For any given input string $x$ in $\Sigma^*$,
a {\em move} of each prover $P_i$ is dictated by an application of its unitary {\em strategies}   $\{U^x_{P_i,j}:\Gamma_{i}\times\Delta_{i}^{\infty}\rightarrow\Gamma_{i}\times\Delta_{i}^{\infty}\}_{j\in\nat^+}$, where each unitary operator $U^x_{P_i,j}$ modifies,  at each step $j$, the contents of the $i$th communication cell and of a {\em finite} segment\footnote{Therefore, at any step, there are only a finite number of tape cells that hold non-blank symbols.} of the $i$th private tape.
Conventionally, we assume that all the provers make their actions simultaneously in a single step without communicating with each other.
As long as it is clear from the context, the above-described protocol of the $k$QIP system $(P_1,P_2,\ldots,P_k,V)$ is also written as $(P_1,P_2,\ldots,P_k,V)$.
In this paper, we define the {\em running time} of a $k$QIP protocol along a computation path $p$ to be the total number of moves made by the verifier and $k$ provers along this particular computation path $p$. We often take the {\em expectation}, over all computation paths $p$, of the running time of the $k$QIP protocol along the computation paths $p$.

One round of {\em interaction} between $V$ and $P_1,\ldots,P_k$ comprises the following three stages: (i) each prover returns a communication symbol to the verifier except at the first step, (ii) with receiving the prover's answer, the verifier applies $\delta_{V}$ and sends the obtained symbols to all the provers, and (iii) the verifier conducts a projection measurement to observe its current inner state to determine whether it is in a halting inner state. After the measurement, only computation paths associated with non-halting inner states continue to the next round.
The $k$QIP protocol $(P_1,\ldots,P_k,V)$ is said to {\em accept} (resp., {\em reject}) {\em $x$ with probability $\gamma$} if the overall probability of entering accepting (resp., rejecting) inner states is exactly $\gamma$. The language recognition criteria are given as follows. Let $a(n)$ and $b(n)$ be any functions from $\nat$ to the real interval $[0,1]$. A language $L$ is said to {\em admit an $(a(n),b(n))$-$k\qip$ system $(P_1,\ldots,P_k,V)$} if the following two conditions hold:
\renewcommand{\labelitemi}{$\circ$}
\begin{itemize}
  \setlength{\topsep}{-2mm}%
  \setlength{\itemsep}{1mm}%
  \setlength{\parskip}{0cm}%
\item {\sf (completeness)} for every string $x\in L$, the $k\qip$ protocol $(P_1,\ldots,P_k,V)$ accepts $x$ with probability at least $a(|x|)$, and

\item {\sf (soundness)} \sloppy for every string $x\not\in L$ and every set of $k$ provers $P^*_1,\ldots,P^*_k$, the $k\qip$ protocol $(P^*_1,\ldots,P^*_k,V)$ rejects\footnote{Note that, as shown by Lipton (see Condon's survey \cite{Con93}) in a classical single-prover model, another choice of soundness property of {\lq\lq}accepting $x$ with probability $< 1-b(|x|)${\rq\rq} makes the IP system significantly more powerful.}
    with probability at least $b(|x|)$.
\end{itemize}
For simplicity, we say that {\em the $k$QIP system $(P_1,\ldots,P_k,V)$ recognizes $L$ with error probability at most $\epsilon(n)$}
if $L$ admits an $(a(n),b(n))$-$k\qip$ system with the condition that $\epsilon(n)\geq\max\{1-a(n),1-b(n)\}$ for all lengths $n\in\nat$.
In the rest of this paper, we will often treat $a(n)$, $b(n)$, and $\epsilon(n)$ as constant functions (or just constants).

The notation $k\qip_{K}(2qfa)$ expresses the collection of all languages recognized by $(1-\epsilon,1-\epsilon)$-$k\qip$ systems with
2qfa-verifiers for certain error bound $\epsilon\in[0,1/2)$ (those systems are generally referred to as {\em bounded-error systems}),
where the verifier's transition amplitudes are drawn from a set $K$ (called an {\em amplitude set}) of certain complex numbers.
When $K=\complex$, we often drop the subscript $K$. Since we are interested only in bounded-error systems, in the rest of this paper, we shall often omit the reference to the error probability $\epsilon$ of the systems for the sake of conciseness.

By expanding the above notation further, let $k\qip(\pair{restrictions})$ denote  the class of languages that admit $k$QIP systems with ``restrictions'' specified by $\pair{restrictions}$.  Of all possible such restrictions, we intend to   consider the following typical ones: $\pair{1qfa}$ (measure-many one-way quantum finite automata), $\pair{2qfa}$ (measure-many two-way quantum finite automata), and $\pair{\polytime}$ (expected polynomial running time).  As an example, the notation $k\qip(2qfa,\polytime)$ indicates that we use $k$-prover QIP systems that have 2qfa-verifiers who run in expected polynomial-time, communicating through the communication cells with $k$ quantum provers.
Moreover, we write\footnote{In this paper, we use a more conventional notation $\qmip(2qfa)$ rather than $\mathrm{MQIP}(2qfa)$ although the notation $\mathrm{MQIP}(2qfa)$ seems more natural because it directly expands the existing notation $k\qip(2qfa)$ for $k$-prover QIP systems.} $\qmip_{K}(\pair{restrictions})$ for the union of all $k\qip_{K}(\pair{restrictions})$ for any constant $k\in\nat^{+}$.

\subsection{Power of QMIP Systems}\label{sec:bounds}

Naturally, we can anticipate that the power of multiple-prover QIP systems  with weak verifiers  significantly exceeds the power of single-prover QIP systems, because each verifier may shrewdly exploit more than one prover to prevent the others from cheating the verifier.
As a major contribution of this paper, we shall prove that the language-recognition power of 2qfa-verifier QMIP systems matches the power of nondeterministic exponential-time Turing machines if the associated verifiers halt in expected polynomial time. In the case where running time is ignored, the power of 2qfa-verifier QMIP systems equals the power of Turing machines. Furthermore, when  the verifiers are restricted to 1qfa's, we give two reasonable bounds on the power of the corresponding QMIP systems.

Those results are succinctly summarized in the following theorem.

\begin{theorem}\label{two-quantum-prover}{\rm (main theorem)}
\renewcommand{\labelitemi}{$\circ$}
\begin{enumerate}\vs{-1}
  \setlength{\topsep}{-2mm}%
  \setlength{\itemsep}{1mm}%
  \setlength{\parskip}{0cm}%
\item $\mathrm{CFL}\subseteq \qmip_{\ptcomplex}(1qfa) \subseteq \mathrm{NE}$.

\item $\qmip_{\ptcomplex}(2qfa) = \mathrm{RE}$.

\item $\qmip_{\ptcomplex}(2qfa,\polytime) = \nexp$.
\end{enumerate}
\end{theorem}

This theorem exhibits a stark contrast with the earlier results in \cite{NY04b,NY09} that $\qip(1qfa) = \mathrm{REG}$ and $\qip_{\ptcomplex}(2qfa,\polytime)\subseteq \np$. Since $\np\neq\nexp$ and $\mathrm{REG}\neq \mathrm{CFL}$, we immediately obtain two (anticipated) separations between QIP systems and QMIP systems.

\begin{corollary}
\renewcommand{\labelitemi}{$\circ$}
\begin{enumerate}
  \setlength{\topsep}{-2mm}%
  \setlength{\itemsep}{1mm}%
  \setlength{\parskip}{0cm}%
\item $\qip(1qfa)\neq \qmip(1qfa)$.

\item $\qip_{\ptcomplex}(2qfa,\polytime)\neq \qmip_{\ptcomplex}(2qfa,\polytime)$.
\end{enumerate}
\end{corollary}

Throughout the rest of this paper, we shall verify our main theorem, Theorem \ref{two-quantum-prover}.

\section{Proofs of the Main Theorem}

Let us give the proof of Theorem \ref{two-quantum-prover}. For the reader's convenience, since the three assertions
in the main theorem are quite similar in nature, we shall describe in detail
the proof of the third assertion $\qmip_{\ptcomplex}(2qfa,\polytime)=\nexp$,
and we shall give only a brief explanation of how to modify its proof to obtain the proofs of the other assertions.
To improve the readability of the proof,
we shall split the proof into two technical parts,
in which we shall employ quite different techniques. Hereafter, we shall prove the containment $\qmip_{\ptcomplex}(2qfa,\polytime) \subseteq \nexp$ in Section \ref{sec:appox-power}, whereas Section \ref{sec:simulation-MIP} will discuss the opposite containment $\nexp \subseteq \qmip_{\ptcomplex}(2qfa,\polytime)$ (actually, $\nexp \subseteq \twoqip_{\algebraic}(2qfa,\polytime)$).

\subsection{Approximation of the Power of QMIP Protocols}\label{sec:appox-power}

This subsection aims at proving that, for any fixed index $k\geq2$,   $k\qip_{\ptcomplex}(2qfa,\polytime)\subseteq \nexp$. Our proof is based on a direct simulation of a QMIP system on an appropriate NTM.
This is a natural  extension of the proof of the containment $\qip_{\ptcomplex}(2qfa,\polytime)\subseteq \np$, given by Nishimura and Yamakami \cite{NY04b,NY09}, based on their single-prover model.

The simulation of a $k$QIP system on an NTM requires us to trace not only every move of a given verifier but also any move of $k$ provers. Naturally, there are two obstacles to cope with. Whereas a 1qfa-verifier halts within $n+1$ steps (where $n$ is any input size), a 2qfa-verifier may possibly produce computation paths of ``arbitrary'' lengths. In addition, provers may use an ``unlimited''  amount of space in his private tape. For our simulation, it is therefore necessary to limit the behaviors of the verifier as well as $k$ provers. Fortunately, the bounded-error requirement of the $k$QIP system allows us to prune long computation paths of the verifier. As
Kobayashi and Matsumoto \cite{KM03} demonstrated, there exists a way to curtail the prover's private work space usage without altering the acceptance probability of the original $k$QIP system. We shall state this result as a key proposition, Proposition \ref{QIP-bounded-system}.

To describe this proposition, we first need to introduce a notion of
{\em resource-bounded $k$QIP system}. Given a function $s$ from $\nat\times\nat$ to $\nat$, a prover is called {\em $s$-space bounded} (or $s(n,i)$-space bounded, for emphasizing $(n,i)$) if he uses only the first $s(n,i)$ cells of his private tape at any step $i$ and on any input $x$ of length $n$ \cite{NY04b,NY09}.
When $i$ is irrelevant, we dare to drop $i$ from the notation $s(n,i)$.
A {\em $(t(n),s(n))$-bounded $k$QIP system} $(P_1,\ldots,P_k,V)$ is a variant of a QMIP system that is obtained from a $k$QIP system by forcing the corresponding $k$QIP protocol to terminate after $t(n)$ steps against $s(n)$-space bounded provers. When the prover's space is not bounded, we use the notation ``$\infty$''. For convenience, after the {\em forced} termination, any non-halting inner state that the verifier currently takes is interpreted  as  the outcome of ``I don't know.''

\begin{proposition}\label{QIP-bounded-system}
Let $k\geq2$. Every language in $k\qip_{\ptcomplex}(2qfa,\polytime)$
admits an  $(n^{O(1)},n^{O(1)})$-bounded $k$QIP system.
Moreover, in the case of 1qfa-verifiers, any language in $k\qip_{\ptcomplex}(1qfa)$ admits an  $(O(n),O(n))$-bounded $k$QIP system.
\end{proposition}

In comparison with this proposition, we note from \cite{NY04b,NY09} that the size of the prover's private work space in a {\em single-prover system} with 2qfa-verifiers is bounded by $O(\log{n})$, which is independent of the number of steps taken by the prover. The desired bounds of given  verifiers and provers, stated in the proposition, can be directly obtained from the following two technical lemmas.

\begin{lemma}\label{run-time}
Let $k\geq1$. Let $L$ be a language recognized by a $k$-prover QIP protocol $(P_1,\ldots,P_k,V)$ with 2qfa-verifier $V$ running in expected polynomial time and with $k$ provers $P_1,\ldots,P_k$. There exist a polynomial $p$ and a QIP protocol with a 2qfa-verifier $V'$ and $k$ provers that recognizes $L$ with bounded error probability, when forcing the verifier to halt exactly after $p(n)$ steps for each input size $n$.
\end{lemma}

\begin{lemma}\label{private-tape-size}
Let $k\geq1$. Let $(P_1,\ldots,P_k,V)$ be any $(t(n),\infty)$-bounded $k$QIP protocol with a 2qfa verifier $V$ and $k$ provers $P_1,\ldots,P_k$. Let $\Gamma_i$ be the communication alphabet of prover $P_i$, where $i\in[1,k]_{\integer}$. There exist $2t(n)\ceilings{\log{\hat{\gamma}}}$-space bounded provers $P'_1,\ldots,P'_k$ for which a $k$QIP protocol $(P'_1,\ldots,P'_k,V)$ is $(t(n),2t(n)\ceilings{\log{\hat{\gamma}}})$-bounded and also has the same acceptance/rejection probability as $(P_1,\ldots,P_k,V)$, where $\hat{\gamma} = \max_{1\leq i\leq k}\{|\Gamma_i|\}$.
\end{lemma}

Lemma \ref{run-time} is obtained simply by analyzing the success probability of a computation tree of a given QMIP protocol on a given input. Lemma \ref{private-tape-size} is an adaptation of the aforementioned result of Kobayashi and
Matsumoto \cite{KM03}. Nishimura and Yamakami \cite{NY04b,NY09} gave the proof of this lemma for the single-prover case (\ie $k=1$).
The proof for the multiple-prover case is similar and rather
straightforward.

To complete our simulation of a $k$QIP system on an NTM, we need another key proposition that establishes a simulation of a $(t(n),s(n))$-bounded $k$QIP system on an NTM. Earlier, Nishimura and Yamakami \cite{NY04b} demonstrated how to simulate single-prover resource-bounded QIP systems on multi-tape NTMs. With a slight  modification, we can make their simulation procedure work for multiple-prover systems. Therefore, we obtain the following proposition.

\begin{proposition}\label{bounded-QIP-on-NTM}
Every language recognized by a certain a $\ptcomplex$-amplitude $(t(n),s(n))$-bounded $k$QIP system with a 2qfa-verifier belongs to $\mathrm{NTIME}(n^{O(1)}t(n)2^{O(s(n))}(\log{t(n)})^{O(1)})$.
\end{proposition}

The next lemma, given in \cite{NY04a}, relates to the approximation of a given unitary operator by a certain quantum circuit of modest size. Here, we
fix an appropriate universal set  of quantum gates, with $\tilde{\complex}$-amplitudes,
consisting of the Controlled-NOT gate and a finite number of single-qubit
gates that generate a dense subset
of SU(2) with their inverse. For our convenience, we abbreviate
$(\log{n})^d$ as $\log^d{n}$ for each constant $d$ in $\nat^{+}$.

\begin{lemma}\label{qustring-gate}{\cite{NY04a}}\hs{1}
For any sufficiently large number $d\in\nat^{+}$, any $d$-qubit unitary
operator $U_d$, and any real number $\epsilon>0$, there
exists a quantum circuit
$C$ of size at most
$2^{3d}\log^3{(1/\epsilon)}$ acting on $d$ qubits satisfying that
$\|U_{C}-U_d\|<\epsilon$, where $U_{C}$
is the unitary operator computing $C$ and $\|A\|$ denotes the norm $\sup_{\qubit{\phi}\neq0}\|A\qubit{\phi}\|/\|\qubit{\phi}\|$.
\end{lemma}

With the use of this lemma, let us verify the containment of  $k\qip_{\ptcomplex}(2qfa,\polytime)\subseteq \nexp$ for any fixed index $k\geq2$. To simplify our proof description, we shall show only the essential case of $k=2$. Its generalization to the case of $k\geq3$ is rather straightforward and is left to the reader.

Let $L$ be any language in $2\qip_{\ptcomplex}(2qfa,\polytime)$. By Proposition \ref{QIP-bounded-system}, there exists a $(p(n),p(n))$-bounded 2QIP system $(P_1,P_2,V)$ with 2qfa-verifier $V$ running in expected polynomial time with error probability at most $\epsilon$, where $p$ is a certain nonnegative polynomial and $\varepsilon$ is a constant lying in the real interval $[0,1/2)$.  We want to simulate this system on a multi-tape model of NTM in  exponential time.

Consider the following simulation algorithm. Let $x$ be any string of length $n$. Prepare additional three work tapes, each of which simulates the behavior of one of $P_1$, $P_2$, and $V$ within time polynomial in $n$.
Notice that all transition amplitudes of $V$ are approximable to within $2^{-n}$ in polynomial time. By Lemma \ref{qustring-gate}, we can replace $P_1$ and $P_2$ by quantum circuits $C_1$ and $C_2$, respectively, of size polynomial in $n$. Let $n$ be an input size. Choose nondeterministically two quantum circuits $C_1$ and $C_2$ of polynomial size. Let $C_V$ be a quantum circuit that simulates $V$ approximately to within $2^{-n}$.
Step by step, we simulate each move of the QMIP protocol using $C_1$, $C_2$, and $C_V$. Finally, we simulate $C_1$, $C_2$, and $C_V$ approximately to within $2^{-n}$ on certain NTMs. Thus, it is possible to make the total simulation error of the QMIP protocol on a certain NTM bounded by $O(2^{-n})$.

Since the error bound of the original proof system is at most $\varepsilon$, the total error caused by the above NTM is at most $\varepsilon+O(2^{-n})$. Since $n$ is sufficiently large, $L$ must be recognized by this NTM. Note that the running time of the NTM is exponential in $n$. This yields $L\in\nexp$, as requested. Therefore, it immediately follows that $2\qip_{\ptcomplex}(2qfa,\polytime) \subseteq \nexp$.


In the case of 1qfa-verifiers, a similar argument proves that, with the help of the second part of Proposition \ref{QIP-bounded-system},
the above simulation of a given QMIP system requires only time $2^{O(n)}$;
thus, the containment $k\qip_{\ptcomplex}(1qfa)\subseteq \mathrm{NE}$ follows immediately.
The remaining containment, $k\qip_{\ptcomplex}(2qfa)\subseteq \mathrm{RE}$, can be obtained accordingly by removing the time bound of 2qfa-verifiers from the above argument.

\subsection{Simulation of Classical MIP Protocols}\label{sec:simulation-MIP}

We shall argue the remaining containments
$\nexp \subseteq \twoqip_{\algebraic}(2qfa,\polytime)$, $\mathrm{RE} \subseteq \twoqip_{\algebraic}(2qfa)$, and $\cfl\subseteq \twoqip_{\algebraic}(1qfa)$.

In the early 1990s, Feige and Shamir \cite{FS92} demonstrated that $\twoip(2pfa)=\mathrm{RE}$ and $\twoip(2pfa,\polytime)=\nexp$.
A key idea of their proofs is that a certain 2pfa verifier can simulate a verifier $V$ who runs a polynomial-time probabilistic Turing machine (or PTM, in short), by forcing two provers to hold the content of a work tape of the verifier.
This simulation can be guaranteed by splitting the whole tape content into two pieces and each prover manipulates only one piece. By the definition, the provers are disallowed to communicate with each other, and therefore they take no chance of collaborating to recover the original tape content.

Although Feige and Shamir did not discuss 1pfa-verifier 2IP systems,
the above argument helps us prove that $\mathrm{CFL}$ is included in $\twoip(1pfa)$. In fact, we can show a slightly stronger statement as shown below than $\cfl\subseteq \twoip(1pfa)$.

\begin{proposition}
Let $L$ be any language that is recognized by an NTM having a read-once input tape (where a head always moves rightward) and a two-way work tape, which runs for exactly $n$ steps, where $n$ is the input size. This language $L$ admits
a 2IP system with 1pfa verifier (whose tape head always moves).
As a special case, $\cfl\subseteq \twoip(1pfa)$ holds.
\end{proposition}

\sloppy 
To achieve our goal of this section, it suffices to prove that $\twoip(1pfa)\subseteq \twoqip_{\algebraic}(1qfa)$,
$\twoip(2pfa)\subseteq \twoqip_{\algebraic}(2qfa)$, and $\twoip(2pfa,\polytime)\subseteq \twoqip_{\algebraic}(2qfa,\polytime)$.

In what follows, we shall present the proof of the last containment:  $\twoip(2pfa,\polytime) \subseteq \twoqip_{\algebraic}(2qfa,\polytime)$. To prove this containment, if a verifier $V$ is equipped with a mechanism of {\em observing} messages sent from multiple provers, then $V$ can project the quantum information received from them onto classical information and this procedure seems to give a simple and clean proof of the desired containment. Nevertheless, since our verifier performs no measurement on any communication symbols, another way is definitely needed to implement a similar procedure.
To make our proof clean, we shall take the following two steps.
(i) We shall simulate any 2pfa-verifier 2IP system by an appropriately chosen 3QIP system whose third prover always removes every message sent from the verifier to its private tape (we call such a  prover an {\em eraser} in our discussion). (ii) We shall reduce such a 3QIP system with an eraser into another ``equivalent'' 2QIP system.

We call a 2qfa-verifier $V$ {\em restrictive} if its transition function $\delta_{V}$ is made up only of the form:
\[
\delta_{V}(q,\sigma,\sigma_1,\sigma_2,\sigma_3)
=  \alpha_1 \qubit{p_1,d_1,\tau_{1,1},\tau_{2,1},\tau_{3,1}}
 +  \alpha_2 \qubit{p_2,d_2,\tau_{1,2},\tau_{2,2},\tau_{3,2}},
\]
where $q,p_1,p_2\in Q$, $\sigma\in \Sigma$, $d_1,d_2\in\{0,\pm1\}$,  $\sigma_i,\tau_{i,1},\tau_{i,2}\in\Gamma_{i}$ for each index $i\in\{1,2,3\}$, and $\alpha_1,\alpha_2\in\complex$ with $|\alpha_1|^2+|\alpha_2|^2=1$.
In the subsequent argument, we shall fix a specific amplitude set  $K_0=\{0,\pm1,\pm1/\sqrt{2}\}$, which is obviously a subset of $\algebraic$.
The aforementioned two steps are formally stated as follows.

\begin{lemma}\label{classical-quantum}
Let $L$ be any language over alphabet $\Sigma$. Let $a,b\in(0,1]$. If $L$ has an $(a,b)$-2IP system with an expected-polynomial-time 2pfa-verifier (resp., a 1pfa-verifier), then $L$ has a $K_0$-amplitude $(a,b)$-3QIP system  whose verifier is a restrictive, expected-polynomial-time 2qfa (resp., a restrictive 1qfa) and the third prover acts as an eraser.
\end{lemma}

\begin{lemma}\label{two-four-qmip}
Let $a,b\in(0,1]$. If a language $L$ over alphabet $\Sigma$ has an $\algebraic$-amplitude $(a,b)$-3QIP system whose verifier is a restrictive,  expected-polynomial-time 2qfa (resp., a restrictive 1qfa) and the third prover is an eraser, then there exists an $\algebraic$-amplitude $(a,b)$-2QIP system, whose verifier is a 2qfa running in expected polynomial time (resp., a 1qfa)  for $L$.
\end{lemma}

The desired containment $\twoip(2pfa,\polytime)\subseteq \twoqip_{\algebraic}(2qfa,\polytime)$ easily follows from Lemmas \ref{classical-quantum} and \ref{two-four-qmip} in the following fashion.
Initially, let us take an arbitrary language $L$ in $\twoip(2pfa,\polytime)$. Applying Lemma
\ref{classical-quantum} to $L$, we obtain a $K_0$-amplitude  3QIP system whose verifier is a restrictive, expected-polynomial-time 2qfa and whose third prover acts as an eraser. Subsequently,  apply Lemma \ref{two-four-qmip} to this system and obtain an $\algebraic$-amplitude 2QIP system for $L$, whose verifier is also a 2qfa running in expected polynomial time. Therefore, $L$ belongs to $\twoqip_{\algebraic}(2qfa,\polytime)$, as requested.

The rest of this section is devoted to verifying the aforementioned  two  lemmas. To prove Lemma \ref{classical-quantum}, since our target system is classical, our proof nails down to a simulation of a 2pfa-verifier $V$ by an appropriately chosen 2qfa-verifier $V'$. A direct simulation of $V$  encounters a difficulty because it may make ``irreversible'' moves; nonetheless, an {\em indirect} simulation is possible if we can store all the information on the moves of the 2pfa-verifier. For this purpose, we utilize an eraser who keeps such information in his private tape without interfering with the verifier and the other provers.

More precisely, whereas we set our honest quantum provers $P'_1$ and $P'_2$ to be the same as the given classical provers, we make the third honest quantum prover $P'_3$ send only the blank symbol $\#$ to a verifier, say, $V'$ and remove every symbol sent from the verifier directly into its private tape with no modification of the sent symbol as well as the symbols that have been written already on the private tape.
Our 2qfa-verifier $V'$ simulates $V$ step by step except for the treatment of $P'_3$.
Against the third prover $P'_3$, if  $V$ in inner state $q$ scans symbol $\sigma$ and two messages $(\sigma_1,\sigma_2)$ sent from the other two provers and is about to send them back new messages $(\tau_1,\tau_2)$,
then  $V'$ sends the information on the choice of $V$ (\ie $(q,\sigma,\sigma_1,\sigma_2,\tau_1,\tau_2)$) to $P'_3$
in order to ensure the reversibility of the behavior of $V'$. In addition, whenever $P'_3$ intentionally sends back any non-$\#$ symbol, $V'$ must detect  the dishonesty of $P'_3$ and immediately reject $x$. This extra detection procedure forces $P'_3$ to keep being honest.

Let us begin with the formal proof of Lemma \ref{classical-quantum}.

\begin{proofof}{Lemma \ref{classical-quantum}}
Assume that $L$ admits an $(a,b)$-2IP system $(P_1,P_2,V)$ with 2pfa-verifier $V$  running in expected polynomial time.  For simplicity, we assume that $V$ always tosses a {\em fair coin} (causing $V$'s computation to form a binary tree). We want to simulate the corresponding 2IP protocol $(P_1,P_2,V)$ by a certain $K_0$-amplitude 3QIP protocol $(P'_1,P'_2,P'_3,V')$ with expected polynomial-time 2qfa-verifier $V'$ and eraser $P'_3$.
To make our proof readable, we assume without loss of generality that $P_1$ and $P_2$ use only {\em reversible strategies}. This can be done by preserving on their private tapes all information necessary to make their strategies reversible.

\sloppy
Formally, we define $P'_1$ and $P'_2$ to be $P_1$ and $P_2$, respectively.
The transition function $\delta_{V'}$ of $V'$ is defined from $\delta_{V}$
of $V$ as follows. If the transition function $\delta_{V}$ has the form
$\delta_{V}(q,\sigma,\sigma_1,\sigma_2) = \{(p_1,d_1,\tau_{1,1},\tau_{2,1}),(p_2,d_2,\tau_{1,2},\tau_{2,2})\}$ with $(p_1,d_1,\tau_{1,1},\tau_{2,1})\neq (p_2,d_2,\tau_{1,2},\tau_{2,2})$,
 then we define
\[
\delta_{V'}(q,\sigma,\sigma_1,\sigma_2,\#)
= \frac{1}{\sqrt{2}}
\bigqubit{p_1,d_1,\tau_{1,1},\tau_{2,1},\track{q\sigma\sigma_1\sigma_2}{\tau_{1,1}\tau_{2,1}}} + \frac{1}{\sqrt{2}}
\bigqubit{p_2,d_2,\tau_{1,2},\tau_{2,2},\track{q\sigma\sigma_1\sigma_2}{\tau_{1,2}\tau_{2,2}}}.
\]
In addition, to force the third prover to play as an eraser, we include
the following rule: for any non-blank communication symbol $\xi$,
\[
\delta_{V'}(q,\sigma,\sigma_1,\sigma_2,\xi)
= \bigqubit{q_{rej},+1,\#,\#,\track{q\sigma\sigma_1\sigma_2}{\xi}}.
\]
In any other cases, we set $\delta_{V'}$'s value arbitrarily as long as $V'$ preserves the unitarity.

It still remains to prove that the new system $(P'_1,P'_2,P'_3,V')$ is an expected-polynomial-time 2qfa-verifier $(a,b)$-3QIP system for
$L$. Let $x$ be any input string of length $n$. When $x\in L$, the protocol $(P'_1,P'_2,P'_3,V')$ obviously accepts $x$ with probability at least $a$ because so does $(P_1,P_2,V)$.

Next, assume that $x\not\in L$. Let $P^*_1$, $P^*_2$, and $P^*_3$ be three (adversarial) quantum provers who maximize the acceptance probability of $V'$. It is therefore possible to assume that the protocol $(P^*_1,P^*_2,P^*_3,V')$ is never aborted by $V'$ in the middle of its computation. A key observation is that, since $P^*_3$ keeps  the history of all moves of $V'$ in his private tape without returning any past information to the verifier, all the computation paths of $V'$ do not interfere with one another.
To show the soundness property of the protocol $(P^*_1,P^*_2,P^*_3,V')$, we wish to define two classical provers $\tilde{P}_1$ and $\tilde{P}_2$
in the following manner.
First, let us define $\tilde{P}_1$. This is done by specifying his $i$th strategy, namely, a map from $(\sigma,y)\in \Gamma_1\times\Delta_1^{\infty}$ to a certain symbol, say,  $\tau_{\sigma,y,i}$. In a computation of $(P^*_1,P^*_2,P^*_3,V')$, let
$\qubit{\sigma}\qubit{y}$ be a content of the first communication cell and the first private tape, obtained just after
step $i-1$. Assume that $U^x_{P^*_1,i}$ satisfies  $U^x_{P^*_1,i}\qubit{\sigma}\qubit{y}= \sum_{\tau\in\Gamma_1} \alpha_{\sigma,y,\tau}\qubit{\tau}\qubit{\phi_{\sigma,y,\tau}}$, where $\qubit{\phi_{\sigma,y,\tau}}$ is a certain unit-norm quantum state with $\alpha_{\sigma,y,\tau}\in\complex$. Note that $\sum_{\tau\in\Gamma_1}|\alpha_{\sigma,y,\tau}|^2=1$.
Let $p_{rej}(\sigma,y,\tau)$ denote the total  rejection  probability of the computation subtree with root $(\sigma,y)$, which is generated by traversing
the computation paths associated with $\tau$.
Let us choose the lexicographically-first symbol $\tau'$ that makes  $p_{rej}(\sigma,y,\tau')$ minimal. Finally, we set this $\tau'$ as the desired symbol $\tau_{\sigma,y,i}$. It then follows that
\[
\sum_{\tau\in\Gamma_1}|\alpha_{\sigma,y,\tau}|^2p_{rej}(\sigma,y,\tau) \geq \sum_{\tau\in\Gamma_1}|\alpha_{\sigma,y,\tau}|^2\cdot
\min_{\tau'\in\Gamma_1}\{p_{rej}(\sigma,y,\tau')\} = p_{rej}(\sigma,y,\tau_{\sigma,y,i}).
\]
Likewise, we define $\tilde{P}_2$. Employing the mathematical induction, we can verify that the rejection probability of the protocol $(\tilde{P}_1,\tilde{P}_2,V)$ does not exceed that
of $(P^*_1,P^*_2,P^*_3,V')$. Since
$(\tilde{P}_1,\tilde{P}_2,V)$
rejects $x$ with probability at least $b$, the protocol $(P^*_1,P^*_2,P^*_3,V')$ must reject $x$ with probability at least $b$.

Therefore, $(P'_1,P'_2,P'_3,V')$. is indeed an $(a,b)$-3QIP system recognizing $L$.
\end{proofof}

Finally, we shall give the proof of Lemma \ref{two-four-qmip}. What we need here
is to  simulate three provers by two provers. It is rather easy to simulate three {\em honest} provers by two {\em honest} provers in such a way that, for instance, one prover simulates two original provers. However, this simple strategy fails when all the three provers simultaneously try to cheat a weak verifier. Therefore, we need to implement a method of detecting any wrongdoing of the provers. Our possible strategy is that, when the verifier wants to send a symbol $\tau_1$ to the first prover and a symbol $\track{\tau_2}{\tau_3}$ to the second prover, he randomly generates a symbol $r$ and embeds it into those  two symbols to prevent the provers from tampering the information. Now, the verifier sends symbols $\track{\tau_1}{r}$ and $\track{\tau_2}{r\oplus \tau_3}$ to the provers, provided that $\tau_3$ and $r$ are expressed in binary and of the same length, where the notation $r\oplus s$ denotes the {\em bitwise XOR} of $r$ and $s$. Such a use of the random variable $r$ completely hides $\tau_3$ from the provers. For simplicity of our proof given below, we also force the verifier to reject immediately whenever he receives any symbol of the form $\track{\sigma}{s}$ with $s\neq\#$.

\begin{proofof}{Lemma \ref{two-four-qmip}}
Let $(P_1,P_2,P_3,V)$ be any restrictive, $\algebraic$-amplitude  $(a,b)$-3QIP system with expected-polynomial-time 2qfa-verifier $V$ and eraser $P_3$ for two constants $a,b\in(0,1]$.
We wish to define another expected-polynomial-time 2qfa-verifier $\algebraic$-amplitude 2QIP system $(P'_1,P'_2,V')$, which simulates  $(P_1,P_2,P_3,V)$ with the same error probability.
For simplicity, we assume that $\Gamma_1=\Gamma_2=\Gamma_3$ and we concisely  write $\Gamma$ for them. We also assume that every symbol in $\Gamma$ is expressed in binary and of the same length. Thus, a transition function $\delta_{V}$ of $V$ can be viewed as a map from $Q\times \Gamma^3$ to $\algebraic^{Q\times \{0,\pm1\}\times\Gamma^3}$.

We begin with the definition of
the desired 2qfa-verifier $V'$. For this purpose, it suffices to define $\delta_{V'}$.
Since $V$ is restrictive, let us firstly assume that $V$ has a transition of the form:
\[
\delta_{V}(q,\sigma,\sigma_1,\sigma_2,\#) =
\alpha_1\qubit{p_1,d_1,\tau_{1,1},\tau_{2,1},\tau_{3,1}} +
\alpha_2\qubit{p_2,d_2,\tau_{1,2},\tau_{2,2},\tau_{3,2}},
\]
where $|\alpha_1|^2+|\alpha_2|^2=1$ and $(p_1,d_1,\tau_{1,1},\tau_{2,1},\tau_{3,1}) \neq (p_2,d_2,\tau_{1,2},\tau_{2,2},\tau_{3,2})$.
We then define the corresponding transition of $V'$ as
\[
\delta_{V'}\left(q,\sigma,\track{\sigma_1}{\#},\track{\sigma_2}{\#}\right)
= \sum_{r\in\Gamma}\frac{1}{\sqrt{|\Gamma|}} \left( \sum_{j=1,2} \alpha_j \bigqubit{p_j,d_j,\track{\tau_{1,j}}{r},\track{\tau_{2,j}}{r\oplus \tau_{3,j}}} \right).
\]
Since  $(p_1,d_1,\tau_{1,1},\tau_{2,1},\tau_{3,1}) \neq (p_2,d_2,\tau_{1,2},\tau_{2,2},\tau_{3,2})$,  it holds that,  for all pairs  $r,r'\in \Gamma$,  $(p_1,d_1,\track{\tau_{1,1}}{r},\track{\tau_{2,1}}{r\oplus \tau_{3,1}})
\neq
(p_2,d_2,\track{\tau_{1,2}}{r'},\track{\tau_{2,2}}{r'\oplus \tau_{3,2}})$.  This property leads to the equality of
\begin{eqnarray*}
\lefteqn{ \delta_{V'}\left(q,\sigma,\track{\sigma_1}{\#},\track{\sigma_2}{\#}\right) \cdot \delta_{V'}\left(q',\sigma',\track{\sigma'_1}{\#},\track{\sigma'_2}{\#}\right) } \hs{5} \\
&=& \sum_{i,j=1,2} \frac{1}{|\Gamma|} \sum_{r\in\Gamma} \measure{ p_i,d_i,\track{\tau_{1,i}}{r},\track{\tau_{2,i}}{r\oplus \tau_{3,i}}}{ p'_j,d'_j,\track{\tau'_{1,j}}{r},\track{\tau'_{2,j}}{r\oplus \tau'_{3,j}}} \\
&=&
\frac{1}{|\Gamma|} \sum_{r\in\Gamma} \sum_{i,j=1,2}
\measure{p_i,d_i,\tau_{1,i},\tau_{2,i},\tau_{3,i}}{ p'_j,d'_j,\tau'_{1,j},\tau'_{2,j},\tau'_{3,j}} \\
&=& \sum_{r\in \Gamma} \frac{1}{|\Gamma|}
\left( \delta_{V}\left(q,\sigma,\sigma_1,\sigma_2,\#\right) \cdot \delta_{V}\left(q',\sigma',\sigma'_1,\sigma'_2,\#\right) \right)
\;\;=\;\; \delta_{V}\left(q,\sigma,\sigma_1,\sigma_2,\#\right) \cdot \delta_{V}\left(q',\sigma',\sigma'_1,\sigma'_2,\#\right).
\end{eqnarray*}
Secondly, if $\delta_{V}(q,\sigma,\sigma_1,\sigma_2,\#)$ is of the form $
\alpha \qubit{p,d,\tau_1,\tau_2,\tau_3}$, then we define
\[
\delta_{V'}\left(q,\sigma,\track{\sigma_1}{\#},\track{\sigma_2}{\#}\right)
= \sum_{r\in\Gamma}\frac{\alpha}{\sqrt{|\Gamma|}} \bigqubit{p,d,\track{\tau_1}{r},\track{\tau_2}{r\oplus \tau_3}}.
\]
A simple observation proves that the inner product of any two transitions of $V'$ preserves that of $V$. From this fact, the unitarity of $\delta_{V'}$ follows immediately.
To complete the definition of $\delta_{V'}$, whenever either $s_1\neq \#$ or $s_2\neq \#$ holds,  we further define
\[
\delta_{V'}\left(q,\sigma,\track{\sigma_1}{s_1},\track{\sigma_2}{s_2}\right)
= \bigqubit{q_{rej},+1,\track{q\sigma_1}{s_1},\track{\sigma\sigma_2}{s_2}}.
\]

Finally, let us define two honest provers $P'_1$ and $P'_2$ so that, for every index $i\in\{1,2\}$, $P'_i$ simulates $P_i$ together with transferring any symbol in the lower track, which is received from the verifier, into an unused (blank) area of his private tape without altering the symbol itself.

Now, we shall show that $(P'_1,P'_2,V')$ is an $(a,b)$-2QIP system for $L$.
Let $x$ be any instance of $L$ and assume that $x\in L$; namely,  the 3QIP protocol $(P_1,P_2,P_3,V)$ accepts $x$ with probability at least $a$. Since $P_3$ is an eraser, the 2QIP protocol $(P'_1,P'_2,V')$ can simulate $(P_1,P_2,P_3,V)$ without introducing any additional error; therefore,   $(P'_1,P'_2,V')$ accepts $x$ with the same probability as $(P_1,P_2,P_3,V)$ does.

Next, assuming that $x\not\in L$, we consider two arbitrary provers $P^*_1$ and $P^*_2$.
Toward a contradiction, we assume that the protocol $(P^*_1,P^*_2,V')$ rejects $x$ with probability less than $b$.
Without loss of generality, we can assume that $P^*_1$ and $P^*_2$ never send illegal symbols (\ie $\track{\sigma}{s}$ with $s\neq\#$) since,  otherwise, the verifier immediately rejects $x$, causing to increase the rejection probability.
Let us define three new provers $\tilde{P}^*_1$, $\tilde{P}^*_2$, and $\tilde{P}^*_3$.
Suppose that a strategy $U_{P^*_1,i}$ of $P^*_1$ at step $i$ transforms  $\left(\sum_{r\in\Gamma}|\Gamma|^{-1/2} \qubit{\track{\sigma}{r}}\right)\otimes \qubit{\phi}$ to $\sum_{\tau\in\Gamma}\qubit{\track{\tau}{\#}}\qubit{\psi_{\tau}}$. In this case, we define a strategy  $U_{\tilde{P}^*_1,i}$ to be a map from $\qubit{\sigma} \qubit{\phi}$ to $\sum_{\tau\in\Gamma}\qubit{\tau}\qubit{\psi_{\tau}}$.
How can we construct this $U_{\tilde{P}^*_1,i}$? This construction
can be done by combining the following series of operations: firstly generate $\sum_{r\in\Gamma}|\Gamma|^{-1/2}\qubit{\track{\sigma}{r}}$ from $\qubit{\sigma}$, secondly apply $U_{P^*_1,i}$, and finally transform $\qubit{\track{\tau}{\#}}$ to $\qubit{\tau}$. Note that $\tilde{P}^*_1$ keeps the same content of the private tape of $P^*_1$.
In a similar manner, we can define $\tilde{P}^*_2$.
Finally, let $\tilde{P}^*_3$ be an eraser that always maps $\qubit{\sigma}\qubit{\phi}$ to $\qubit{\#}(\qubit{\sigma}\qubit{\phi})$ where $\ket{\sigma}\ket{\phi}$ is now written onto the prover's private work tape.
By the above definition, the rejection probability of the protocol $(\tilde{P}^*_1,\tilde{P}^*_2,\tilde{P}^*_3,V)$ is at most that of $(P^*_1,P^*_2,V')$, and thus it must be less than $b$. This immediately leads to a contradiction because $(\tilde{P}^*_1,\tilde{P}^*_2,\tilde{P}^*_3,V)$ should reject $x$ with probability at least $b$.

Therefore, $(P'_1,P'_2,V')$ is an $(a,b)$-2QIP system recognizing $L$.
This completes the proof of the lemma.
\end{proofof}


\let\oldbibliography\thebibliography
\renewcommand{\thebibliography}[1]{%
  \oldbibliography{#1}%
  \setlength{\itemsep}{0pt}%
}
\bibliographystyle{plain}

\end{document}